\documentclass[aps,pra,twocolumn,groupedaddress,showpacs,showkeys]{revtex4}

\usepackage{graphicx}
\begin{document}

\newcommand{\be}{\begin{equation}}
\newcommand{\ee}{\end{equation}}

\title{General theory  of electromagnetic fluctuations   near  a homogeneous
surface, in terms of its reflection amplitudes\\}

\author{Giuseppe Bimonte}
\author{Enrico Santamato}
\affiliation{%
Dipartimento di Scienze Fisiche Universit\`{a} di Napoli Federico
II Complesso Universitario MSA Via Cintia
I-80126 Napoli Italy;\\ INFN, Sezione di Napoli, Napoli, ITALY\\
}

\date{\today}

\begin{abstract}
We derive  new general expressions for the fluctuating
electromagnetic field outside a homogeneous material surface. The
analysis is based on general results from the thermodynamics of
irreversible processes, and requires no consideration of the
material interior, as it only  uses knowledge of the reflection
amplitudes for its surface. Therefore, our results are valid for
all homogeneous surfaces, including layered systems and
metamaterials, at all temperatures. In particular, we obtain new
formulae for the near-field region, which are important for
interpreting the numerous current experiments probing proximity
effects for macroscopic and/or microscopic bodies separated by
small empty gaps. By use of Onsager's reciprocity relations, we
obtain also the general symmetry properties that must be satisfied
by the reflection matrix of any material.

\end{abstract}


\maketitle

\section{Introduction}

The study of electromagnetic (e.m.) fluctuations has received much
attention over the years, because of its fundamental importance in
diverse areas of physics. While the classical problem of
determining the thermal spectrum of e.m. fluctuations in a black
body is at the origin of quantum theory, the concept of zero-point
e.m. quantum fluctuations  leads to new quantum phenomena, like
the Lamb shift, and the existence of van der Waals  forces
between atoms and/or macroscopic bodies. Electromagnetic
fluctuation-driven forces become appreciable in the submicron
range and rapidly increase in the nanometer scale.

It should be observed that while general thermodynamical theorems
(Kirchhoff's law) give much information on the radiation field far
from the body surface, no analogously general results are
available so far, on the problem of characterizing the e.m. field
very close to a surface. A detailed study of the near-field region
is of great importance  in numerous physical phenomena that depend
on proximity effects between macroscopic and/or microscopic bodies
separated by small empty gaps. Important examples of such
phenomena are the Casimir effect \cite{casimir}, radiative heat
transfer between closely space bodies at different temperatures
\cite{rytov,polder,persson}, and vacuum friction \cite{pendry}.
Non contact interactions mediated by fluctuations of the e.m.
field are now finding also  exciting applications to
nanotechnology \cite{capasso}, to Bose-Einsten condensates
\cite{antezza}, and to ultracold atoms \cite{carusotto}.

The theory of e.m. fluctuations and thermal radiation was
developed long ago by Rytov \cite{rytov}. In this theory, the
physical origin of e.m. fluctuations  resides in the microscopic
fluctuating currents, which are present in the interior of any
absorbing medium, according to the fluctuation-dissipation
theorem. These random currents,
when introduced into Maxwell's equations, constitute the source of
fluctuating e.m. fields, that extend beyond the boundaries of the
medium, partly in the form of propagating waves (PW) and partly as
evanescent waves (EW) localized near the surface. Rytov's method was
used by Lifshitz \cite{lifs} in his famous work of van-der Waals
attractions between macroscopic solid bodies, which constitutes the
basis of all current studies of the Casimir effect between real media.
The same ideas were also applied by Polder and van Hove \cite {polder}
to investigate the problem of radiative heat transfer between closely
spaced dielectric bodies. While appealing for its conceptual
simplicity, a drawback of Rytov's approach is that it requires
detailed knowledge of the e.m. fields in the body's interior. In
particular, the theory requires as inputs the dielectric tensor
$\epsilon_{ij}(\omega,{\bf x}, {\bf x}')$ and/or the magnetic
permittivity $\mu_{ij}(\omega,{\bf x}, {\bf x}')$ of the material,
together with the boundary conditions satisfied by the electric and
magnetic fields at the interface. Therefore, with this approach, the
problem arises  of understanding to what extent the final expressions
for the fluctuating e.m. field outside the body depend on the
assumptions made for the permittivities and the chosen boundary
conditions, which
%
is a very delicate issue in the case, say, of spatially non-local
materials.

The spirit of our approach is radically different, as we never
consider the e.m. field or the microscopic fluctuating currents in
the interior of the material body. Therefore, we do not make any
use of either the constitutive equations for the material, or the
boundary conditions for the e.m. field across the interface. We
regard the surface as fully characterized in terms of a set of
reflection amplitudes $r_{\lambda \mu}=r_{\lambda \mu}(\omega,{\bf
k}_{\perp})$, depending on the frequency $\omega$ and the
projection ${\bf k}_{\perp}$ of the wave-vector onto the surface.
These amplitudes are considered by us as data, obtained either by
experiments or by an independent computation, which is of no
concern to us. We shall see that knowledge of the reflection
amplitudes is sufficient to determine, in a completely general
way,  the correlators for the e.m. field {\it outside} the body.
%
The possibility of expressing the correlators of the e.m. field
outside a material surface in terms of its reflection amplitudes is
not
entirely
new. It is clearly suggested by Lifshitz formula \cite{lifs}, as in
its final form it expresses the interaction between two nearby
parallel dielectric slabs in terms of their respective reflection
amplitudes. The general validity of Lifshitz formula, once expressed
in terms of reflection amplitudes, has been recognized by several
authors, and it has been recently demonstrated using the scattering
approach \cite{jaekel,esquivel}. The idea that field correlators can
be given in terms of reflection amplitudes was also utilized in the
study of radiative heat transfer, including evanescent contributions,
in Ref. \cite{persson}. The distinctive feature of our approach is
that it relies on general results from the thermodynamics of
irreversible processes, and requires no consideration about the field
in the interior of the body. The process of reflection is considered
as a macroscopic irreversible process, entirely described in terms of
the reflection amplitudes 
at the surface. Therefore, our results are of a very general nature,
and they apply to arbitrary homogeneous surfaces at all temperatures,
thus
providing a generalization of previously available results, that
includes the case of non-diagonal reflection matrices, i.e. mixing TE
and TM modes. While for PW we recover the well known Kirchhoff's
formula, Eq. (\ref{spw}) below, we obtain a new general and simple
formula for the EW contribution, Eq. (\ref{sew}), which constitutes
the
%
main result of this paper.  By exploiting Onsager's reciprocity
thermodynamical relations, we obtain also new general symmetry
properties that must be satisfied by the reflection matrix of any
material.

\section{The e.m. field outside a homogeneous surface}

In this Section we derive the expression for the correlators of
the e.m. field outside a homogeneous surface. The procedure we
follow   is a generalization of the method used in Ref.
\cite{callen} to obtain the Planck radiation law. In that paper
the authors considered an electric dipole in thermal equilibrium
with the radiation field of a black body. Apart from the
intra-dipole binding force, the dipole is subjected to two forces
of e.m. origin: one of them is a systematic damping force,
representing the reaction to the radiation emitted by the dipole
itself, and the other is a random force due to the fluctuating
electric field of the black body (to linear order, magnetic
interactions can be neglected). The damping force is computed
using Maxwell's Equations, and then one can use the general
fluctuation-dissipation theorem to obtain
%
the power spectrum of the fluctuating e.m. field, which gives the
Planck's radiation law. In our case, we are interested in determining
how the presence of a surface influences the radiation field in the
vacuum region bounded by the surface (for simplicity we limit our
analysis to flat homogeneous surfaces), and we shall see that the
problem can be solved in full generality by means of a similar
procedure, but considering, instead of one, {\it two} electric dipoles
placed   on a  plane parallel to surface.

To be definite, let us suppose that the body occupies the $z>0$
half-space, such that its surface is at $z=0$. By invariance under
translations in the $x,y$ plane, the fluctuating electric field in
the vacuum region outside the body (i.e. in the $z<0$ half-space)
can be written in  general as:
\begin{widetext}\be {\bf E}^{(\rm fluc)}({\bf x}_{\perp},z)=\int_0^{\infty} \frac{d
\omega}{2 \pi}\int \frac{d \,{\bf k}_{\perp}}{(2 \pi)^2}\;({\bf
E}^{(\infty)}(z;\omega,{\bf k}_{\perp})+{\bf
E}^{(S)}(z;\omega,{\bf k}_{\perp}))\;e^{i({\bf k}_{\perp} {\ensuremath\cdot}
{\bf x}_{\perp}-\omega t)}\;+ {\rm c.c.}\;,\label{etot} \ee
\end{widetext}
where
\begin{widetext}
\begin{eqnarray}
{\bf E}^{(\infty)}(z;\omega,{\bf
k}_{\perp})=\left[a_s^{(\infty)}(\omega,{\bf k}_{\perp})\,{\bf
e}_{\perp}\right. \!\! &+& \!\!\left. a_p^{(\infty)}(\omega,{\bf
k}_{\perp})\frac{c}{\omega}({\bf k {\times} e}_{\perp}) \right]e^{i
k_z z}\; +   \,\nonumber \\
  +\left\{ a_s^{(\infty)}(\omega,{\bf k}_{\perp}) \left[ \frac{}{}r_{ss}{\bf
e}_{\perp}\right. \right.\!\!&+&\!\!\left. \left.
r_{p\,s}\frac{c}{\omega}({\bf k}^{(r)} {\bf{\times}
e}_{\perp})\right]+a_p^{(\infty)}(\omega,{\bf k}_{\perp})\left[
r_{pp}\frac{c}{\omega}({\bf k}^{(r)} {\bf{\times}
e}_{\perp})+r_{sp}\,{\bf e}_{\perp}\right] \right\}e^{-i k_z
z}\;,\label{einfty}\\
 {\bf E}^{(S)}(z;\omega,{\bf
k}_{\perp})=\left[a_s^{(S)}(\omega,{\bf k}_{\perp})\,{\bf
e}_{\perp}\right.\!&+&\!\left.a_p^{(S)}(\omega,{\bf
k}_{\perp})\frac{c}{\omega}({\bf k^{(r)} {\times} e}_{\perp})
\right]e^{-i k_z z}\;\label{es}.
\end{eqnarray}
\end{widetext}
In the above Equations, ${\bf k}={\bf k}_{\perp}+k_z {\bf{\hat
z}}$ and ${\bf k}^{(r)}={\bf k}_{\perp}-k_z {\bf{\hat z}}$ denote,
respectively, the wave-vectors for fields propagating towards and
away from the surface, $k_z=\sqrt{\omega^2/c^2-k_{\perp}^2}$, and
we take the square root such that $k_z' \ge 0$, $k_z'' \ge 0$
(with prime and double prime denoting real and imaginary parts,
respectively). We recall that real $k_z$ correspond to PW waves,
while imaginary $k_z$ describe EW. Moreover, ${\bf x}_{\perp}=x
\,{\bf \hat{x}}+y \,{\bf \hat{y}}$, ${\bf e}_{\perp}={\bf{\hat z}
{\times} {\hat k}}_{\perp}$, the subscripts $s$ and $p$ denote $TE$
and $TM$ polarizations, respectively, and $r_{\lambda
\mu}=r_{\lambda \mu}(\omega,{\bf k}_{\perp})$ are the reflection
amplitudes for an incident field with polarization $\mu$ to be
reflected as a field with polarization $\lambda$. The field ${\bf
E}^{(\infty)}$ represents an incoming field from $z=-\infty$, that
gets reflected by the surface, while we can think of ${\bf
E}^{(S)}$ as the field radiated by the surface. Since the
expansion for the magnetic field is obtained from Eqs.
(\ref{etot}-\ref{es}) by integrating Maxwell's equations (${\bf
B}=-c \int dt \,{\bf \nabla {\times} E}$), the fluctuating e.m.
field is fully described by the correlators among the amplitudes
$a^{(\infty/S)}_{s/p}(\omega,{\bf k}_{\perp})$. We let
$a^{(\alpha)}\;,\;\alpha=\infty,S$ a column vector with elements
$(a^{(\alpha)}_{s},a^{(\alpha)}_p)$, and $a^{(\alpha)\dagger}$ a
row vector with elements $(a^{(\alpha)*}_{s},a^{(\alpha)*}_p)$.
From invariance under time translations, and by homogeneity in the
$x,y$ plane, the correlators must be of the form:
\begin{widetext} \be \langle a^{(\alpha)} (\omega,{\bf
k}_{\perp}),a^{(\beta)\dagger}(\omega',{\bf k}_{\perp}') \rangle=
(2 \pi)^3\; C^{(\alpha \beta)} (\omega,{\bf k}_{\perp})\;
\delta(\omega - \omega')\,\delta^{(2)}({\bf k}_{\perp}-{\bf
k}_{\perp}')\;,\ee\end{widetext} with all other correlators
vanishing. Therefore, the problem reduces to computing the four
matrices $C^{(\alpha \beta)}(\omega,{\bf k}_{\perp})$. To do this,
we imagine placing in the radiation field two electric dipoles. We
let $q^{(A)}$ and ${\bf \xi}^{(A)}$, $A=1,2$ their respective
charges and displacements, and we assume that they are placed at
arbitrary points ${\bf x}_{\perp}^{(1)}$ and ${\bf
x}_{\perp}^{(2)}$ on a plane $z=w<0$ in the empty region below the
surface. For our purposes, it is convenient to suppose that the
displacements ${\bf \xi}^{(A)}$ of both dipoles are bound to occur
in the $x,y$ plane, ${\bf \xi}^{(A)}={\bf \xi}^{(A)}_{\perp}$. For
sufficiently small $q^{(A)}$ and ${\bf \xi}_{\perp}^{(A)}$, the
Equations of motion for the two oscillators are linear, and have
the following form:
\begin{widetext}\be m^{(A)} \ddot {\bf
\xi}^{(A)}_{\perp}(t)+k^{(A)} {\bf \xi}^{(A)}_{\perp}(t)- q^{(A)}
\sum_{B=1,2}q^{(B)}\int_{-\infty}^t \!\!dt'
\stackrel{\leftrightarrow}{\bf E}_{\perp}^{(\rm
ret)}\!\!(t-t',{\bf x}^{(A)}_{\perp}-{\bf
x}^{(B)}_{\perp},w)\,\dot {\bf \xi}^{(B)}_{\perp}(t')=q^{(A)}{\bf
E}^{\rm (fluc)}_{\perp}(t,{\bf
x}^{(A)}_{\perp},w)\;.\label{lang}\ee
\end{widetext}
In this Equation, $k^{(A)}$ are the elastic constants for the
restoring intra-dipolar forces, while the non-local tensor kernel
$\stackrel{\leftrightarrow}{\bf E}_{\perp}^{(\rm ret)}(t-t',{\bf
x}^{(A)}_{\perp}-{\bf x}^{(B)}_{\perp},w)$ represents projection
onto the $x-y$ plane of the retarded electric field produced by
dipole $B$ at the position of dipole $A$, which produces a
systematic frictional force. Finally, the force on the r.h.s. is a
random term arising from the fluctuating electric field. The
expression of the time Fourier transform of the kernel
$\stackrel{\leftrightarrow}{\bf E}_{\perp}^{(\rm
ret)}\!\!\!(t-t',{\bf x}^{(A)}_{\perp}-{\bf x}^{(B)}_{\perp},w)$
is found by resolving Maxwell Equations in the empty region
outside the body, a task that is  easily accomplished by
exploiting  translational invariance along the surface. Upon
performing a space-Fourier transform of the fields in the
$z$-plane, we obtain:
\begin{widetext}\begin{eqnarray}
\stackrel{\leftrightarrow}{\bf E}_{\perp}^{(\rm
ret)}\!\!(\omega;{\bf x}^{(A)}_{\perp}-{\bf x}^{(B)}_{\perp},w)&=&
\frac{2 \pi}{c} \int \frac{d \,{\bf k}_{\perp}}{(2
\pi)^2}\;\left\{\left[-\frac{\omega}{c k_z}(1+r_{ss}\,e^{-2 i k_z
w})\,{\bf e}_{\perp}-r_{p\,s}e^{-2 i k_z w}\, \hat{\bf
k}_{\perp}\right]\otimes {\bf e}_{\perp}\right.+\nonumber \\
 &+& \left.\left[\frac{c k_z}{\omega}(-1+r_{pp}\,e^{-2 i k_z
w})\,\hat{\bf k}_{\perp}+r_{sp}e^{-2 i k_z w}\,{\bf
e}_{\perp}\right]\otimes \hat{\bf k}_{\perp} \right\}\,e^{i {\bf
k}_{\perp}{\ensuremath\cdot} ({\bf x}^{(A)}_{\perp}-{\bf
x}^{(B)}_{\perp})}\label{eij}
\end{eqnarray}
\end{widetext}
Now, Eq. (\ref{lang}) is a generalized Langevin Equation, of the
type commonly used in the theory of irreversible processes
\cite{kubo}. At thermal equilibrium, it implies a set of general
relations, collectively known as fluctuation-dissipation theorems
\cite{kubo}, involving the frictional forces on one side, and
correlators of random forces or of dynamical variables of the
system, on the other. Of particular interest to us is the
so-called {\it second fluctuation-dissipation theorem}
\cite{kubo}, which in our case reduces to the following relation
between the correlator of the random force and the time Fourier
transform of the damping force:
\begin{widetext}
\be \int_0^{\infty} dt\, e^{i \omega t}\langle {E}_i^{(\rm
fluc)}(t_0+t,{\bf x}^{(A)}_{\perp},w), {E}_j^{(\rm fluc)
}(t_0,{\bf x}^{(B)}_{\perp},w) \rangle =-
F(\omega,T)\,\stackrel{\leftrightarrow}{E}_{ij}^{(\rm
ret)}\!\!(\omega;{\bf x}^{(A)}_{\perp}-{\bf
x}^{(B)}_{\perp},w)\;,\;\;\;\;i,j=x,y\;,\label{sfd}\ee
\end{widetext}
 where $F(\omega,T)$ is the quantity
\be F(\omega,T)\,=\, \frac{\hbar \omega}{2} \,
\coth\left(\frac{\hbar \omega}{2 k_B T}\right)\;,\ee with $T$ the
temperature, and $k_B$ Boltzman's constant. We note that Eq.
(\ref{sfd}) takes account of quantum effects, and includes the
contribution from zero-point fluctuations.  The expressions for
the quantities $C^{(\alpha \beta)}(\omega,{\bf k}_{\perp})$ can be
now determined by inserting Eqs. (\ref{etot}-\ref{es}) into the
l.h.s of Eq. (\ref{sfd}), and requiring that the two members are
equal for arbitrary values of ${\bf x}^{(A)},{\bf x}^{(B)}$ and
$w$. After some tedious but straightforward computations one
obtains for $C^{(\alpha \beta)}(\omega,{\bf k}_{\perp})$ the
following expressions: \be C^{(\infty \infty)}=
F(\omega,T)\,\frac{2 \pi \omega}{c^2}\,{\rm
Re}\left(\frac{1}{k_z}\right)\;,\label{bb}\ee

\be C^{(\infty S)}=C^{(S \infty)}=0\label{cbb}\;.\ee Eq.
(\ref{bb}) coincides with the emission spectrum of a black
surface. and Eqs. (\ref{cbb}) show that this radiation is
uncorrelated with that emitted by the surface. As for $C^{(SS)}$,
when $k_z$ is real, i.e. for PW, we find:  \be C^{(SS)}=\frac{2
\pi \omega}{c^2 k_z}\, F(\omega,T)\,(1-R
\,R^{\dagger})\;,\;\;\;\;{\rm (PW)} \label{spw}\ee while for $k_z$
imaginary, i.e. for EW, we obtain: \be C^{(SS)}=-i\frac{2 \pi
\omega}{c^2 |k_z|}\, F(\omega,T)\,(R
-\,R^{\dagger})\;,\;\;\;\;{\rm (EW)}\label{sew}\ee where $R$ is
the reflection matrix $R_{\lambda \mu}=r_{\lambda \mu}$.  Eqs.
(\ref{spw}) and (\ref{sew}) describe the emission from the
surface.
It should be observed that the contribution from PW, Eq.(\ref{spw}) is
equivalent to Kirchhoff's law, and it is obviously possible to derive
it by simply imposing detailed energy balance between the radiation
flux from the black surface at infinity and the flux radiated by the
surface. The same type of argument, however, is useless to determine
the EW contribution, Eq.(\ref{sew}), because the associated average
flux of energy is zero. Therefore, Eq. (\ref{sew}) should be regarded
as a non-trivial extension of Kirchhoff's formula
%
to the near-field region, and
Eqs. (\ref{spw}) and (\ref{sew})
together
%
provide the complete description of the e.m. field outside the body.
It should be noted that EW fluctuations are zero when the reflection
matrix $R$ is hermitean.

We remark an interesting consequence of the symmetry properties of
the correlation functions, in Eq. (\ref{sfd}). In the absence of
applied external magnetic fields, the correlators on the l.h.s. of
Eq. (\ref{sfd}) are invariant under time reversal, and therefore
the quantities $\stackrel{\leftrightarrow}{E}_{ij}$ on the r.h.s.
satisfy the following symmetry property: \be
\stackrel{\leftrightarrow}{E}_{ij}^{(\rm ret)}\!\!(\omega;{\bf
x}^{(A)}_{\perp}-{\bf
x}^{(B)}_{\perp},w)=\stackrel{\leftrightarrow}{E}_{ji}^{(\rm
ret)}\!\!(\omega;{\bf x}^{(B)}_{\perp}-{\bf
x}^{(A)}_{\perp},w)\;.\label{ons}\ee These relations constitute an
example of {\it Onsager's reciprocity relations}. Upon using Eq.
(\ref{eij}), and recalling that Eq. (\ref{ons}) must be satisfied
for all ${\bf x}^{(A)},{\bf x}^{(B)}$ and $w$, one finds that Eq.
(\ref{ons}) are equivalent to the following symmetry conditions
for the reflection amplitudes: \begin{eqnarray} r_{ss}(\omega,
\vec{k}_{\perp})&=&r_{ss}(\omega, -\vec{k}_{\perp})\;,\\
r_{pp}(\omega,
\vec{k}_{\perp})&=&r_{pp}(\omega, -\vec{k}_{\perp})\;,\\
r_{sp}(\omega, \vec{k}_{\perp})&=&-r_{ps}(\omega,
-\vec{k}_{\perp})\;\label{rps}.
\end{eqnarray}
The above relations can be used a criterion to reject specific
models for the reflection amplitudes. An interesting example of
this sort is provided by the Drude-Born  model for chiral
materials, which is based on the following constitutive Equations
(in the frequency domain): \be {\bf D}= \epsilon \,{\bf E}-f\,
{\bf \nabla {\times} E}\;,\;\;\;\;\;{\bf B}={\bf H}\;. \ee It can be
shown that the reflection amplitudes implied by this model do not
satisfy Eqs. (\ref{rps}), and therefore this model must be
rejected. However, the model for chiral materials proposed by
Fedorov \cite{lakh} \be {\bf D}= \epsilon \,(\,{\bf E} +\beta\,
{\bf \nabla {\times} E})\;,\;\;\;\;\;{\bf B}=\mu \,({\bf H +\beta\,
{\bf \nabla {\times} H}) }\; \ee satisfies the above conditions, and
is therefore viable from the point of view of thermodynamics.
Moreover, it is worth noting that chiral Fedorov's materials, even
if transparent, have a non hermitean reflection matrix $R$.
Proximity effects related to EW near the surface of chiral media
are therefore expected.

\section{Conclusions and discussion}

In conclusion, using general arguments from the thermodynamics of
irreversible processes, we have obtained general formulae for the
fluctuating e.m. fields outside a homogeneous surface. Our results
are valid for all materials, at all temperatures, and provide a
generalization of Kirchoff's law to the near field region. In
particular, they apply to non-diagonal reflection matrices, that
mix TE and TM modes. Moreover, by use of Onsager's reciprocity
relations, we have also obtained the general symmetry properties
that reflection amplitudes must satisfy, in order to be consistent
with the requirements of thermodynamics. The same methods can be
also applied to cavities bounded by homogeneous surfaces, and to
systems out of thermal equilibrium, but we postpone to a separate
paper the discussion of these problems.

\end{document}